\documentclass[showpacs]{revtex4}
\usepackage{indentfirst}
\usepackage[mathscr]{eucal}
\usepackage[dvips]{graphicx}
\usepackage{amsmath}
\def\lsim{\mathrel{\raise2pt\hbox to 8pt{\raise -5pt\hbox{$\sim$}\hss{$<$}}}}

\begin{document}
\title{Solution of the 5D Einstein equations in a dilaton background model}
\author{W. de Paula and T. Frederico}
\affiliation{Dep. de F\'\i sica, Instituto Tecnol\'ogico de Aeron\'autica, CTA, 12228-900 S\~ao Jos\'e dos Campos, Brazil.}
\author{H. Forkel}
\affiliation{Institut fur Physik, Humboldt-Universitat zu Berlin, D-12489 Berlin, Germany}
\author{M. Beyer}
\affiliation{Institute of Physics, University of Rostock, D-18051 Rostock, Germany}
\date{\today}


\begin{abstract}
\noindent We obtain an explicit solution of the $5$d Einstein equations in a dilaton background model. We demonstrate that for each metric ansatz
that only depends on the extra coordinate, it is possible to uniquely determine the dilaton field and its potential consistently with the $5$d
Einstein equation. In this holographic dual model of QCD, conformal symmetry of the Anti-de-Sitter metric near the 4d boundary is broken by a term
that leads to an area law for the Wilson loop. We verify that confinement of the string modes dual to mesons follows from the metric background
and the corresponding dilaton solution of the gravity-dilaton coupled equations. In addition, we show that the meson Regge trajectories constrain
the metric and corresponding dilaton background within the area law requirement. We can also incorporate asymptotic freedom in the gravity
background within the model.

\end{abstract}
\maketitle


The large $N$-limit of gauge theories in $4$d with the dominance of ``planar diagrams" \cite{Thooft}, inspired  the development of the
gauge/gravity duality concept. Maldacena's conjecture \cite{Maldaconj} substantiated the duality concept stating that conformal field theories
(CFT) in particular $N = 4$ $U(N)$ super-Yang-Mills in $4$d is the same or dual to a Type $II B$ string theory in an AdS$_{5} \times$S$_{5}$
space-time. Any operator on the 4d boundary within the gauge/gravity duality has a correspondence to a string field in a higher-dimensional
space-time \cite{Witten,WittenWL,Gubser,Aharony}. In this context, the extension of duality to Quantum Chromodynamics (QCD) could give new
insights and improve our knowledge of the infrared (IR) region, supplying novel analytical tools to study hadronic observables in the
nonperturbative regime of the strong force (see e.g. \cite{Erdmenger}).

The expectation value of the Wilson loop exhibits an area law behavior for a confining gauge theory. The holographic calculation of the Wilson
loop was made by \cite{Malda2,Rey01} and  it was found that for an AdS metric space the potential energy of a quark-antiquark pair goes as $1/L$
instead as $L$ in confining theories.  Later on \cite{Kinar} (see also \cite{Kiritsis}) developed a criterion for the area law behavior for a
given $5$d theory only depending on the bulk geometry.  Therefore, the structure of the bulk space dual to QCD, which has an IR confining scale,
has to be different from AdS$_{5}$. This is realized either by a cutoff in the extra dimension \cite{WittenWL} or by introducing a nontrivial warp
factor that scales as $z^\lambda$ with $\lambda\geq 1$ \cite{Kinar}.

It was shown by \cite{Polchinski} that the conformal invariance of the AdS$_5$ holographic representation of QCD gives the counting rules for the
scaling laws obeyed by  QCD hard scattering amplitudes, while an IR cutoff in the extra dimension, defining a slice of AdS$_{5}$, generates a mass
gap. The cutoff in the extra dimension implies Regge-like trajectories of glueball states \cite{Boschi}, and also in a light hadron mass spectrum
with $m^{2}$ depending quadratically with the principal excitation number $n$ \cite{brodsky}. However,  Regge-like behavior of the spectrum, i.e.,
almost linear trajectories of $m^{2}$  as a function of the spin $(S)$ and excitation number is seen experimentally (see e.g. \cite{Bugg}). Other
approaches to obtain the Regge behavior using a deformed AdS metric were also developed (see e.g. \cite{FBT}).

Ref. \cite{Karch} introduced a dilaton background field, $\Phi(z)\propto z^2$, to induce conformal symmetry breaking in the dual model within an
AdS$_5$ background. They found that $m^2\sim n+S$ for mesons. However, the model is not strictly compatible with the $5$d Einstein equations
without further assumptions. In this respect, a recent paper \cite{Csaki} studied how supergravity can give boundaries to the IR behavior of
AdS/QCD models. It was indicated the necessity of a scalar tachyon besides the dilaton to get the AdS metric as a solution of the $5$d Einstein
(see also \cite{Batell}).

In the present work we use an alternative route. By starting with a deformation of the AdS$_5$ metric we prove that the coupled background dilaton
and $5$d Einstein equations have a consistent solution without the necessity of a scalar tachyon. We also show that the supergravity requirement
of the area law of the Wilson loop expressed by a warp factor scaling with $z^\lambda$ together with the consistent dilaton background confines
the string modes dual to mesons for $\lambda> 1$, giving as well Regge-like trajectories for the spectrum. This holographic model dual to QCD
relies only on two assumptions: i) a dilaton background field and ii) a nonconformal metrics with a boundary condition that implements conformal
symmetry near the 4d brane by the Anti-de-Sitter metric.

We start with the action for five-dimensional gravity coupled to a dilaton:
\begin{equation}
S = \frac{1}{2k^{2}}\int d^{5}x \sqrt{g} \left( -\emph{R} - V(\Phi) + \frac{1}{2}g^{MN}\partial_{M}\Phi\partial_{N}\Phi\right), \label{actiongd}
\end{equation}
where $k$ is the Newton constant in $5$ dimensions and $V(\Phi)$ is the potential of scalar field. Defining the metric
$g_{MN}=e^{-2A(z)}\eta_{MN}$, we find the coupled Einstein equations
\begin{eqnarray}
6A'^{2} - \frac{1}{2}\Phi'^{2} + e^{-2A(z)} V( \Phi ) = 0, \label{einsteinzz}
\\
-3A'^{2} + 3A'' - \frac{1}{2} \Phi'^{2} - e^{-2A(z)} V(\Phi) = 0,
\label{einstein00} \\
\Phi '' - 3 A' \Phi '- e^{-2A(z)}\frac{dV}{d\Phi} = 0. \label{dilatonequation}
\end{eqnarray}
The Einstein equations, (\ref{einsteinzz}) and (\ref{einstein00}), determine the dilaton field directly from the metric model as \cite{dePaula}:
\begin{equation}
\Phi'= \sqrt{3 A'^{2} + 3 A''}~, \label{constrain}
\end{equation}
where we choose the positive sign for the root without loosing generality. Substituting the dilaton field in equation (\ref{einsteinzz}) we obtain
the dilaton potential written as:
\begin{equation}
V(\Phi) = \frac{3 e^{2A}}{2} \left(A''-3 A'^{2}\right) . \label{vz}
\end{equation}

The last step in our derivation is to verify if Eq. (\ref{dilatonequation}) is satisfied by the dilaton from (\ref{constrain}) and by the
potential (\ref{vz}). Using that $ \Phi'\frac{dV}{d\Phi}= \frac{dV}{dz}$ it is straightforward to show that Eq.(\ref{dilatonequation}) is indeed
fulfilled. This proves that the static Einstein equations can be satisfied and the dilaton equation solved for the given model of the metric
background in the extra coordinate without further assumptions. According to \cite{Kiritsis} the IR limit of a general confining metric can be
written as:
\begin{equation}
A(z) = \log \left(z \right)+z^{\lambda}+\dots ~, \label{polinomial_metric}
\end{equation}
where $\lambda$ is a real parameter. Our units are such that the $\mathrm{AdS}_{5}$ radius is unity. The first term reflects the AdS metric that
dominates the UV limit. The second term is the leading one in the IR region, any subleading term is irrelevant for the present discussion of the
Regge trajectory correspondent to the higher excited string states dual to mesons as we will show. The dilaton field obtained by integrating Eq.
(\ref{constrain}) with the boundary condition $\Phi(0)=0$ for $C\left(z\right)=z^\lambda$ is given by:
\begin{eqnarray}
\Phi(z) &=&\frac{\sqrt{3}}{\lambda} \left( (1 +\lambda) \log \left( z^{\lambda/2}\lambda + \sqrt{\lambda+\lambda^2+ z^{\lambda}\lambda^2} \right)+
\right.\nonumber \\ && \left. z^{\lambda/2} \sqrt{\lambda+\lambda^2 + z^{\lambda} \lambda^2}\right)  - \frac{\sqrt{3}}{2\lambda}(1 +\lambda) \log
\left( \lambda+\lambda^2 \right) .
\end{eqnarray}
It gives the following behavior for $z\rightarrow 0$ and $z\rightarrow \infty$, respectively
\begin{eqnarray}
 \Phi(z)\sim c_0
z^{\frac{\lambda}{2}} ~~and~~ \Phi(z)\sim c_\infty z^\lambda , \label{phiz}
\end{eqnarray}
where $c_0=2\sqrt{3 \left(1+1/ \lambda\right)}$ and $c_\infty=\sqrt{3}$.

The dilaton potential for the above confining metric model as a function of the extra dimension is given by:
\begin{eqnarray}
V(\Phi(z))=-\frac32 e^{2z^\lambda}(4+7\lambda z^\lambda+\lambda^2z^\lambda(3 z^\lambda-1)). \label{vphiz}
\end{eqnarray}
We can easily invert the $z$ dependence of Eq. (\ref{vphiz}) in terms of $\Phi$ for the particular limits of $z\rightarrow 0$ and $z\rightarrow
\infty$, using the asymptotic forms written in Eq. (\ref{phiz}). In the limit of $z\rightarrow 0$ we find   the potential:
\begin{eqnarray}
V(\Phi)\sim-6 +\frac{3}{2 c_0^2}(\lambda+1)(\lambda-8)\Phi^2
 \ ,
\end{eqnarray}
and for $z\rightarrow \infty$ we obtain:
\begin{eqnarray}
V(\Phi)\sim-\frac{9}{2 c_\infty^2} \lambda^2\Phi^2 e^{2\Phi/c_\infty} \ ,
\end{eqnarray}
which diverges exponentially reflecting the exponential form of the metric model. We observe that the dilaton potential does not have a lower
bound relying on the coupling to the gravity to stabilize the classical solution. In the UV limit, from the boundary condition for $\Phi(0)=0$ one
recovers the conformal symmetry of the metric with $V(0)=-6$ representing the cosmological constant.

As we have defined the dilaton-metric background of the theory, we are now able to calculate the meson mass spectrum in the spirit of AdS/QCD
duality. To do so, we utilize the AdS/CFT dictionary in the sense that for each operator in the $4$d gauge theory there is a field propagating in
the bulk. For definiteness we follow the notation of ref. \cite{Karch}. The $5$d action for a gauge field $\phi_{M_{1}\dots M_{S}}$ of spin $S$ in
the background is given by
\begin{equation}
I = \frac{1}{2} \int d^{5}x \sqrt{g} e^{-\Phi}\left(\nabla_{N} \phi_{M_{1}\dots M_{S}} \nabla^{N} \phi^{M_{1}\dots M_{S}} \right).
\end{equation}
As in \cite{Karch} and \cite{KatzLewandowski}, we utilize the axial gauge. To this end, we introduce new spin fields $\widetilde{\phi}_{\dots}=
e^{2(S-1)A}\phi_{\dots}$. In terms of this new field, the action is then given by
\begin{eqnarray}
I = \frac{1}{2} \int d^{5}x e^{-5A} e^{-\Phi} e^{-4(S-1)A}e^{2A(S+1)}
\partial_{N}\widetilde{\phi}_{M_{1}\dots M_{S}}
\partial_{N}\widetilde{\phi}_{M_{1}\dots M_{S}} ~.
\label{eqn:newS}
\end{eqnarray}
Using (\ref{eqn:newS}) the equation for the modes $\widetilde{\phi}_{n}$ of the higher spin field $\widetilde{\phi}_{...}$ is derived, viz.
\begin{equation}
\partial_{z}\left(e^{-B}\partial_{z}\tilde{\phi_{n}}\right)
+ m_{n}^{2}e^{-B}\tilde{\phi_{n}} = 0,
\end{equation}
where $B = A (2S-1) + \Phi$. Via the substitution $\tilde{\phi_{n}}= e^{B/2}\psi_{n}$, one obtains a Sturm-Liouville equation
\begin{equation}
\left(-\partial_{z}^{2}+ {\mathcal V}_{eff}(z)\right)\psi_{n} = m_{n}^{2}\psi_{n},
\end{equation}
where the $B$ dependent term in this equation may be interpreted as an effective potential for the string mode, written as ${\mathcal
V}_{eff}(z)=\frac{B'^{2}(z)}{4} -\frac{B''(z)}{2}$. Hence, for each metric $A$ and dilaton field $\Phi$ consistent with the solutions of the
Einstein equations, we obtain a mass spectrum $m_n^2$. Due to the gauge/gravity duality this mass spectrum corresponds to the mesonic resonances
in the $4$d space-time. We can rewrite the effective potential as
\begin{eqnarray}
{\mathcal V}_{eff}(z)= S^{2} A'^{2}+S\left(A'\sqrt{3A'^{2}+3A''}-A'^{2}-A''\right)
 +A'^{2}+\frac{5}{4}A''
\nonumber \\
 -\sqrt{3}~\frac{A''' +4A'A''+
2A'^3}{4\sqrt{A'^{2}+A''}} ~ .
\end{eqnarray}
As an example, let us discuss the  UV and IR properties of the effective potential. For small values of $z$ it can be easily expanded giving:
\begin{eqnarray}
{\mathcal V}_{eff}(z) = \frac{S^2-\frac14}{z^2}+\sqrt{3(\lambda^2+\lambda)}\left(S-\frac{\lambda}{4}\right)z^{\frac{\lambda}{2}-2}
\nonumber \\
+\frac{\lambda}{4}\left(8S^2-S(4\lambda+4)+5\lambda+3\right)z^{\lambda-2}+ \dots , \label{eqn:UVeffP}
\end{eqnarray}
which shows spin dependence in all lower order terms. In the IR limit the metric (\ref{polinomial_metric}) leads to the following effective
potential
\begin{equation}
{\mathcal V}_{eff}(z) \rightarrow \frac{ \lambda^2}{4} (2S-1+\sqrt{3})^{2}\; z^{2\lambda-2}, \label{eqn:effP}
\end{equation}
which presents a discrete spectrum for the normalizable string modes if $\lambda > 1$. It is worthwhile to point out that the analysis of the
effective potential gives a constraint for a confining metric consistent with the one found in the analysis of the Wilson loop (see
\cite{Kiritsis}).

In the case of choosing the dilaton field like $z^2$ for small values of $z$, following the original work (\cite{Karch}), it implies that
$\lambda=4$ for the metric model (\ref{polinomial_metric}). Therefore, using this simple parametrization of the metric the effective IR potential
will grow as $z^6$ deforming strongly the Regge trajectories of the higher radial excitations and spin states. In order to obtain Regge-like
linear trajectories for spin states with high radial excitations, the leading IR term should have $\lambda=2$ instead of $4$, which gives a
harmonic well confinement for large values of $z$. In this case, the meson mass for large radial excitation quantum number ($n>>1$ and $n>>S$) is
$m^2_n\sim 2(2 S-1 +\sqrt{3})~n \ ,$ giving linear Regge trajectories for states with fixed $n$. It is worthwhile to stress that the slope of the
Regge trajectories increases with the excitation number for large $n$. We point out that for large $n$ the AdS mass term in the string action can
be neglected, therefore our result remains valid in this limit.

Of course by modeling the metric with more general forms than the leading power expression in Eq. (\ref{polinomial_metric}) may allow to construct
an effective potential that has a harmonic form for small and large values of $z$, like e.g., $C(z)\sim z^4/(1 + z^2)$. In this way, in our
bottom-up approach of the coupled dilaton-gravity background the fitting of the mesonic spectrum can in principle provide a more detailed form of
the metric of the dual model, as well as the dilaton field.

The dilaton background is constrained by the scalar nature of the corresponding gauge invariant QCD operator. Further constraints would determine
the dilaton UV behavior and the metric model near the physical brane. In this respect we call the attention to the recent work of ref.
\cite{Csaki}, where it was incorporated asymptotic freedom in the dilaton-gravity model described by the action Eq. (\ref{actiongd}) with a warp
factor deviating from AdS near the physical brane with a term like $-(2\log z)^{-1}$. It corresponds to a new UV limit for the meson potential
${\mathcal V}_{eff}(z) = \frac{S^2-1/4}{z^2}+ \frac{\sqrt{3} S}{\sqrt{2} z^2 \log(z)}$.

In conclusion, we showed that it is possible to obtain an explicit solution of the $5$d Einstein equations for a metric ansatz, function of the
extra dimension, in a dilaton background with no necessity of other fields in the model. We have explicitly shown that the metric determines
uniquely the dilaton field and its potential through the 5d Einstein equations that are consistent with the dilaton field equation as well. We
have also shown that the leading IR term in the warp-factor, $z^\lambda$, which allows for conformal symmetry breaking of the Anti-de-Sitter
metric, has to fulfill the condition $\lambda > 1$ to produce a discrete spectrum of the normalizable string modes dual to mesons. The presence of
confinement in our holographic model of QCD has a constraint consistent with the one found imposing an area law for the Wilson loop calculated
within the gauge/gravity duality. In addition, we show that the meson Regge trajectories constrain the IR metric dependence and corresponding
dilaton background consistently with the Wilson loop area law requirement.  We leave for a future work the challenge to describe in more detail
the metrics parameterized according to the experimental light meson spectrum realizing phenomenologically the AdS/QCD duality in a bottom-up
approach.

\begin{acknowledgments}
We thank to DAAD (Deutscher Akademischer Austausch Dienst), CAPES (Coordena\c c\~ao de Aperfei\c coamento de Pessoal de N\'\i vel Superior),
FAPESP (Funda\c c\~ao de Amparo \`a Pesquisa do Estado de S\~ao Paulo) and CNPq (Conselho Nacional de Desenvolvimento Cient\'\i fico e
Tecnol\'ogico) for partial financial support.

\end{acknowledgments}

\end{document}